\documentclass[%
 reprint,
 amsmath,amssymb,
 aps,
]{revtex4-1}
\usepackage{graphicx}
\usepackage{psfrag}
\usepackage{dcolumn}
\usepackage{bm}

\usepackage{siunitx} 
\usepackage{hyperref} 
\usepackage{xcolor}
\hypersetup{
    colorlinks,
    linkcolor={red!50!black},
    citecolor={blue!70!black},
    urlcolor={blue!90!black}
}
\begin{document}

\preprint{APS/123-QED}

\title{Measurement of Ehrlich-Schwoebel barrier contribution to \\the self-organized formation of ordered surface patterns on Ge(001)}
\author{Peco Myint$^1$}
\author{Denise Erb$^2$, Xiaozhi Zhang$^3$, Lutz Wiegart$^4$, Yugang Zhang$^4$, Andrei Fluerasu$^4$, Randall L. Headrick$^3$, Stefan Facsko$^2$} 
\author{Karl F. Ludwig, Jr.$^{1,5}$}
\email{ludwig@bu.edu}

\affiliation{$^1$Division of Materials Science and Engineering, Boston University, Boston, Massachusetts 02215 USA}
\affiliation{$^2$Institute of Ion Beam Physics and Materials Research, Helmholtz-Zentrum Dresden-Rossendorf, Bautzner Landstrasse 400, 01328 Dresden, Germany}
\affiliation{$^3$Department of Physics and Materials Science Program, University of Vermont, Burlington, Vermont 05405 USA}
\affiliation{$^4$National Synchrotron Light Source II, Brookhaven National Lab, Upton, NY 11973}
\affiliation{$^5$Department of Physics, Boston University, Boston, Massachusetts 02215 USA}

\date{\today}
             
\begin{abstract}
{\noindent 
Normal incidence 1 keV Ar$^+$ ion bombardment leads to amorphization and ultrasmoothing of Ge at room temperature, but at elevated temperatures the Ge surface remains crystalline and is unstable to the formation of self-organized nanoscale patterns of ordered pyramid-shaped pits. The physical phenomenon distinguishing the high temperature patterning from room temperature ultrasmoothing is believed to be a surface instability due to the Ehrlich-Schwoebel barrier for diffusing vacancies and adatoms, which is not present on the amorphous material. This real-time GISAXS study compares smoothing of a pre-patterned Ge sample at room temperature with patterning of an initially flat Ge sample at an elevated temperature.  In both experiments, when the nanoscale structures are relatively small in height, the average kinetics can be explained by a linear theory. The linear theory coefficients, indicating surface stability or instability, were extracted for both experiments. A comparison between the two measurements allows estimation of the contribution of the Ehrlich-Schwoebel barrier to the self-organized formation of ordered nanoscale patterns on crystalline Ge surfaces. 

}
\end{abstract}
\maketitle

Elemental semiconductor surfaces bombarded by a broad ion beam at normal incidence are ultrasmoothed \cite{moseler2005ultrasmoothness,madi2009linear} at room temperature. For high polar angle ion bombardment, self-organized nanoscale ripple patterns are observed at room temperature \cite{teichmann2013pattern,perkinson2013nanoscale,anzenberg2012nanoscale, madi2011mass}. However, the surface is amorphized, and the pattern formation is due to instabilities related to the curvature dependence of sputter erosion \cite{sigmund1973mechanism,bradley1988theory}, and to impact-induced mass redistribution \cite{carter1996roughening}, both of which occur together with a stabilizing process or processes such as surface diffusion or surface confined viscous flow \cite{bradley1988theory,carter1996roughening,umbach2001spontaneous}. In contrast, during normal-incidence ion bombardment above the corresponding recrystallation temperature, ordered, faceted patterns are known to form spontaneously along crystal directions on elemental or compound semiconductors \cite{ou2013reverse,ou2015faceted}. The crystalline pattern formation at elevated temperatures is believed to be due mainly due to the surface instability caused by the Ehrlich-Schwoebel (ES) barrier \cite{ou2015faceted}.

In this work, we use real-time experiments with sample temperatures above and below the recrystalliation temperature of Ge to quantitatively measure the ES barrier's contribution to the self-organized pattern formation during 1 keV Ar$^+$ bombardment of Ge. The high temperature sample, which was initially flat, was bombarded at a temperature of 300$^{\circ}$C, leading to self-organized pattern formation of ordered pyramid-shaped pits with four-fold symmetry. On the other hand, normal incidence ion bombardment of a Ge surface at room temperature smooths it.  This hinders X-ray studies by reducing the scattering intensities from the surface.  Therefore, to achieve sufficient scattering intensities, the room temperature experiment was conducted by normal-incidence bombardment of a pre-patterned Ge sample.

For the experiment, 500 $\mu$m thick n-doped (Sb) Ge(001) wafers were cut into 1 $\times$ 1  cm$^2$ pieces and cleaned with acetone, isopropyl alcohol, and methanol.  Samples were firmly affixed to a stage by spot welding Ta strips at the two opposite corners. The water-cooled sample stage base supported a heater which in turn was directly beneath a thermocouple and the sample. For the high temperature bombardment experiment, the sample temperature was kept at 300$^\circ$C. Additionally, the sample stage was electrically isolated except for a wire leading out to an ampere meter in order to measure ion flux. The sample holder was mounted in a custom UHV chamber with mica X-ray windows and a base pressure of 5 $\times$ $10^{-6}$ Torr. For both experiments, samples were bombarded with a broad beam of 1 keV Ar$^+$ ions, which was generated by a 3-cm graphite-grid ion source from Veeco Instruments Inc. placed at 0$^{\circ}$ ion incidence angle ($\theta$). The ion beam flux was measured to be 1 $\times$ 10$^{15}$ ions cm$^{-2}$s$^{-1}$ at the operating chamber pressure of 2 $\times$ $10^{-4}$ Torr. The final fluences were 1.1 $\times$ 10$^{17}$ ions cm$^{-2}$ for the room temperature smoothing and 2.4 $\times$ 10$^{18}$ ions cm$^{-2}$ for the high temperature patterning.

\begin{figure}
\includegraphics[width=3.1 in]{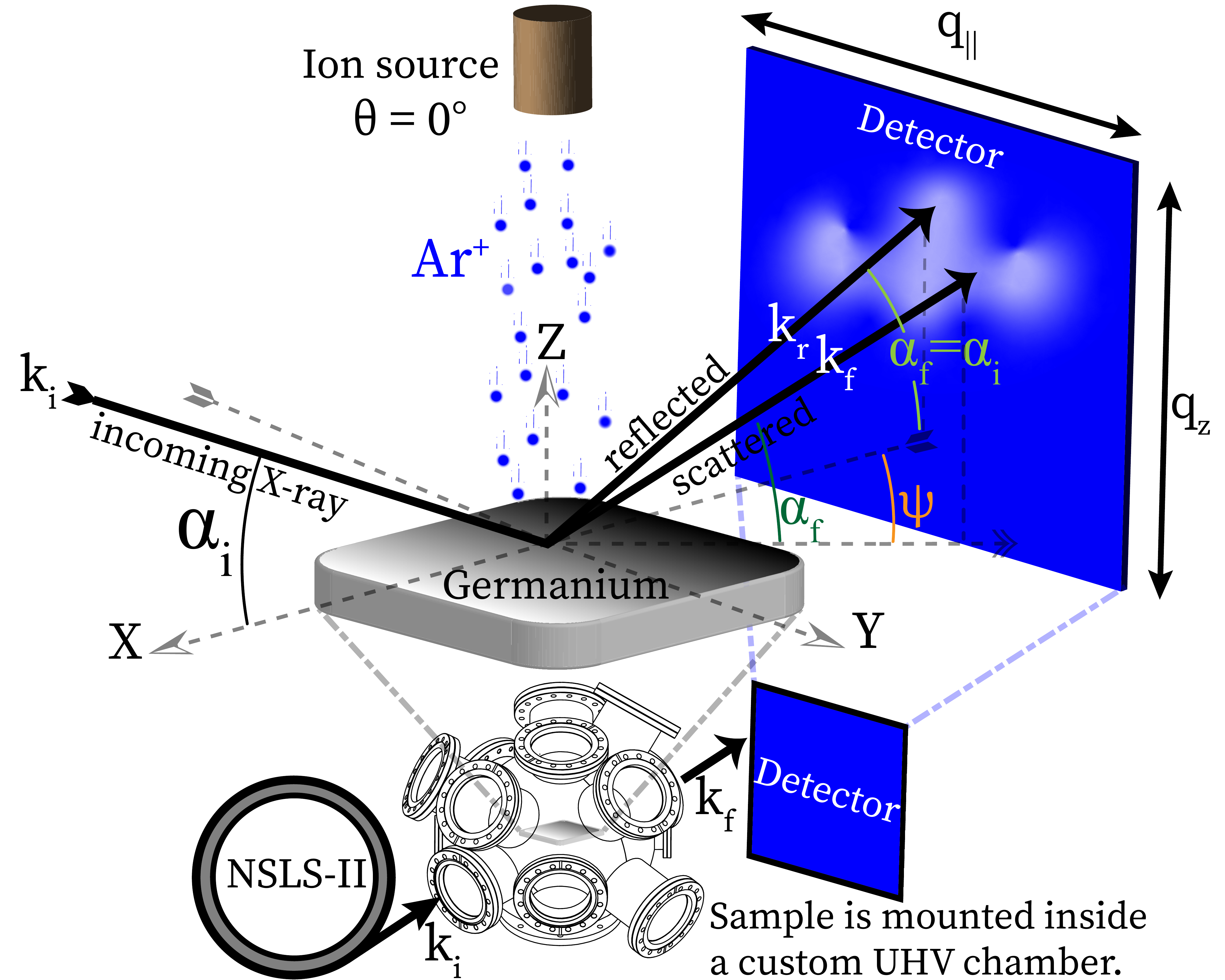}
\caption{A schematic diagram of the GISAXS experiment. The ion source is placed at the polar angle $\theta$ = 0$^{\circ}$, which produces amorphization and ultrasmoothing at low sample temperature and self-organized crystalline patterning at high sample temperature. Ge(001) is positioned so that the X-ray incident angle $\alpha_i$ is slightly above the critical angle of total external reflection. The scattering intensity is recorded as a function of the exit angles $\alpha_i$ and $\psi$ using a 2D detector.}
\label{fig:setup}
\end{figure}

\begin{figure}
    \centering
    \includegraphics[width=3.1 in]{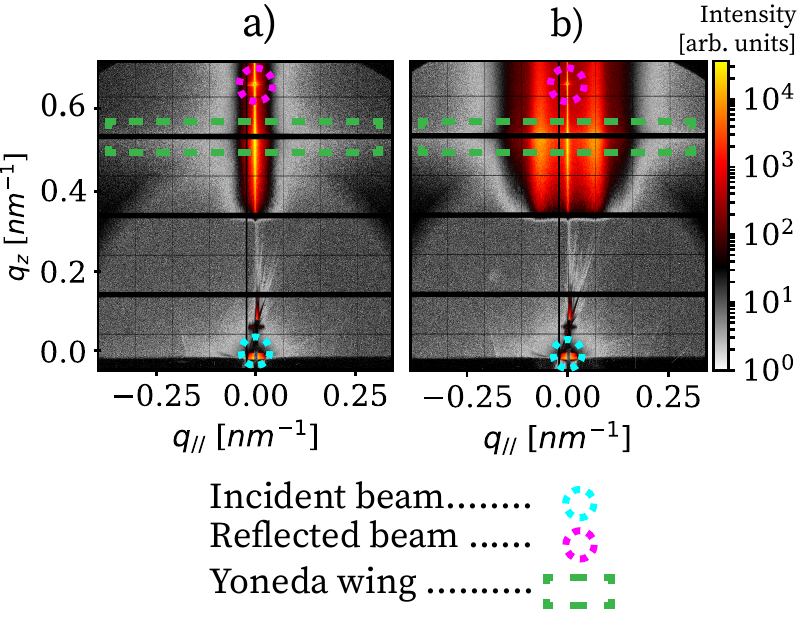}
    \caption{Detector images before and after nanopatterning. The Yoneda wing, indicated by the dashed green box, is the surface-sensitive scattering exiting the sample at the critical angle, $\alpha_c$. a) The scattering pattern of a flat sample before the ion bombardment. b) The scattering pattern after the ion bombardment at 300$^\circ$C. Satellite peaks appeared on each side of the scattering pattern due to the self-organized formation of correlated faceted pit structures.}
    \label{fig:scattering_img}
\end{figure}

Real-time X-ray scattering experiments were performed at the Coherent Hard X-ray (CHX) beamline at the National Synchrotron Light Source-II (NSLS-II) of Brookhaven National Laboratory. The photon energy of 9.65 keV was selected with a flux of approximately 5 $\times$ $10^{11}$ photon s$^{-1}$ and beam dimensions 10 $\times$ 10 $\mu$m$^2$. Experiments used an Eiger-X 4M detector (Dectris) with 75 $\mu$m pixel size, which was located 10.3 m from the sample. The incident X-ray angle $\alpha_i$ was 0.39$^{\circ}$, which is slightly above the critical angle of total external reflection for Germanium of 0.25$^{\circ}$. The scattered intensity was recorded as a function of the exit angle $\alpha_f$  and $\psi$ using the 2D detector, as shown in Fig. \ref{fig:setup}. The change in X-ray wavevector $\mathbf{q}$ can be calculated from those angles:
\begin{equation}
\mathbf{q} = \mathbf{k_f}-\mathbf{k_i} = 
\begin{pmatrix}
  q_x  \\
  q_y  \\
  q_z 
 \end{pmatrix} = \frac{2\pi}{\lambda}
 \begin{pmatrix}
  \cos (\alpha_f) \cos (\psi)- \cos (\alpha_i)  \\
  \cos (\alpha_f) \sin (\psi)  \\
  \sin (\alpha_i)+ \sin (\alpha_f) 
 \end{pmatrix}
 \label{equ:wavenumber_conversion}
\end{equation} 
Since $q_x$ is small, the horizontal component $q_{||}$ (parallel to the surface) can be approximated as simply $q_y$ and the vertical component as $q_z$ (perpendicular to the surface). In the analysis of this paper, we will primarily be interested in the scattering along the Yoneda wing, which is particularly sensitive to surface structure. Example scattering patterns recorded on the area detector are shown in Fig. \ref{fig:scattering_img}.

\begin{figure}
    \centering
    \includegraphics{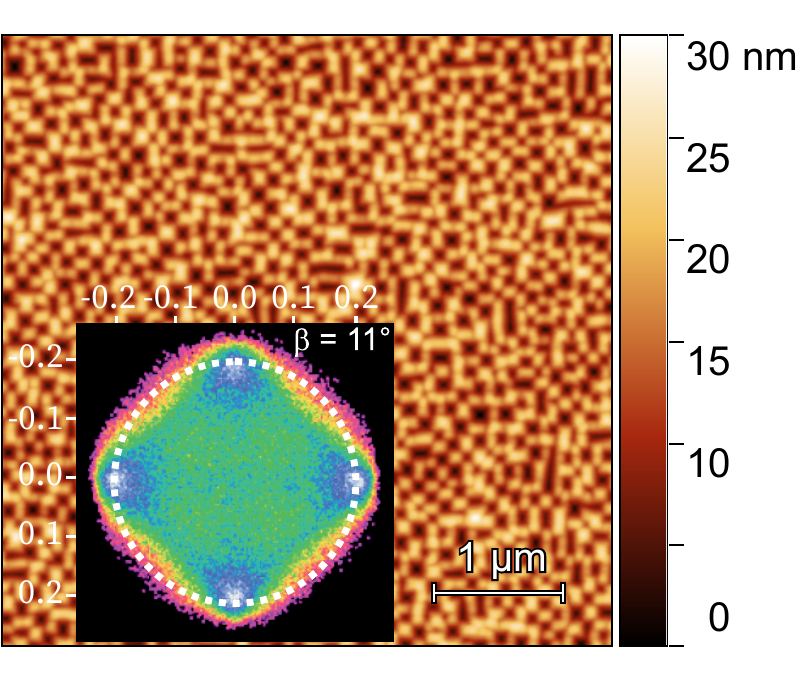}
    \caption{\textit{Post facto} AFM image of Ge(100) surface showing checkerboard patterns of ordered faceted pit structures with fourfold symmetry which formed spontaneously due to 1 keV Ar$^+$ normal incidence bombardment at 300$^\circ$C. The inset shows the distribution of facet slopes, which were calculated through local plane fitting. Four distinct peaks with slope values $\sim \pm$0.2 indicate that the four sides of the pyramid-shaped pit nanostructure have well-defined facet angles of $\sim$ 11$^\circ$.}
    \label{fig:AFM}
\end{figure}

A \textit{post facto} AFM image confirms the formation of ordered faceted patterns on the high temperature sample as seen in Fig. \ref{fig:AFM}. To explain the patterning phenomenon, Ou \textit{et al.} used the following continuum equation \cite{ou2015faceted} in their simulations, which well reproduced the experimental observations of ordered pyramidal pit patterns on Ge: 

\begin{eqnarray}
\frac{\partial h(x,y,t)}{\partial t} &&= - v_0 - \nu \nabla^2h-\nabla\bm{j}_{\mathrm{ion}}-\nabla\bm{j}_{\mathrm{diff}}+ \eta\left(x,y,t\right),\nonumber\\\bm{j}_{\mathrm{diff}} &&= \kappa\nabla(\nabla^2 h)+\sigma\nabla(\nabla h)^2 + \epsilon \left[
\begin{matrix}
h_x (1-\delta h_x^2) \\
h_y (1-\delta h_y^2) \\
\end{matrix}
\right]
\label{equ:Oucontinuum}
\end{eqnarray}
where $v_0$ is the constant erosion rate of the flat Ge surface, $\nu \nabla^2 h $ is the curvature dependent sputter rate, $\bm{j}_{\mathrm{ion}}$ is the surface current resulting from the ballistic mass redistribution, and $\bm{j}_{\mathrm{diff}}$ represents the surface current due to diffusion. $\kappa$ is the isotropic Herring-Mullins (HM) surface diffusion.
$\sigma$ is the non-linear ‘conserved Kadar-Parisi-
Zhang’ coefficient. $\bm{j}_{\mathrm{diff}}$, which has the ES barrier coefficient $\epsilon$ and the parameter $\delta$ (the angle of the facets are given by $\theta = \pm \mathrm{arctan}(\sqrt{1/\delta})$), represents the anisotropic current that incorporates both the anisotropy of the ES barrier itself and step edge diffusion. The last term $\eta\left(x,y,t\right)$ is stochastic noise.

When slopes are small, such as in the early stages of patterning of an initially-flat sample or in the slightly later stages of smoothing an initially patterned surface, the nonlinear terms of Eq. \ref{equ:Oucontinuum} can be neglected and the equation reduces to a linear form, which is then Fourier transformed \cite{bradley1988theory, madi2011mass}:
\begin{equation}
\frac{\partial\tilde{h}\left(\mathbf{q},t\right)}{\partial t}=R\left(\mathbf{q}\right)\tilde{h}\left(\mathbf{q},t\right)+\tilde{\eta}\left(\mathbf{q},t\right)
\label{eq: dispersion-general-form}
\end{equation}
where $\tilde{h}\left(\mathbf{q},t\right)$ is the Fourier transform of the surface height $h\left(x,y,t\right)$, $R\left(\mathbf{q}\right)$ is  the \emph{amplification factor} or \emph{dispersion relation}, and $\tilde{\eta}\left(\mathbf{q},t\right)$ is the Fourier transform of a stochastic noise.  The amplification factor can be determined experimentally by measuring the $\mathbf{q}$-averaged height-height structure factor evolution \cite{madi2011mass,norris2017distinguishing}:

\begin{eqnarray}
\label{equ:hhstructure-factor}
I(\mathbf{q},t)&&= \left\langle h(\mathbf{q},t) \, h^*(\mathbf{q},t)\right\rangle\nonumber\\&& =\left(I_0(\mathbf{q})+\frac{n}{2R(\mathbf{q})}\right)e^{2R(\mathbf{q})t}-\frac{n}{2R(\mathbf{q})}
\end{eqnarray}
where $n$ is the magnitude of the stochastic noise: $\left\langle \eta\left(\mathbf{r},t\right) \eta\left(\mathbf{r^\prime},t\right) \right\rangle = n \, \delta(\mathbf{r}-\mathbf{r^\prime})\delta(t-t^\prime)$. The amplification factor differentiates surface stability or instability; a positive $R(\mathbf{q})$ drives exponential amplification of modes of wavevector $\mathbf{q}$ resulting in surface instability, while a negative $R(\mathbf{q})$ damps fluctuations and stabilizes modes of wavevector $\mathbf{q}$. 

\begin{figure}
    \centering
    \includegraphics[width=3.3 in]{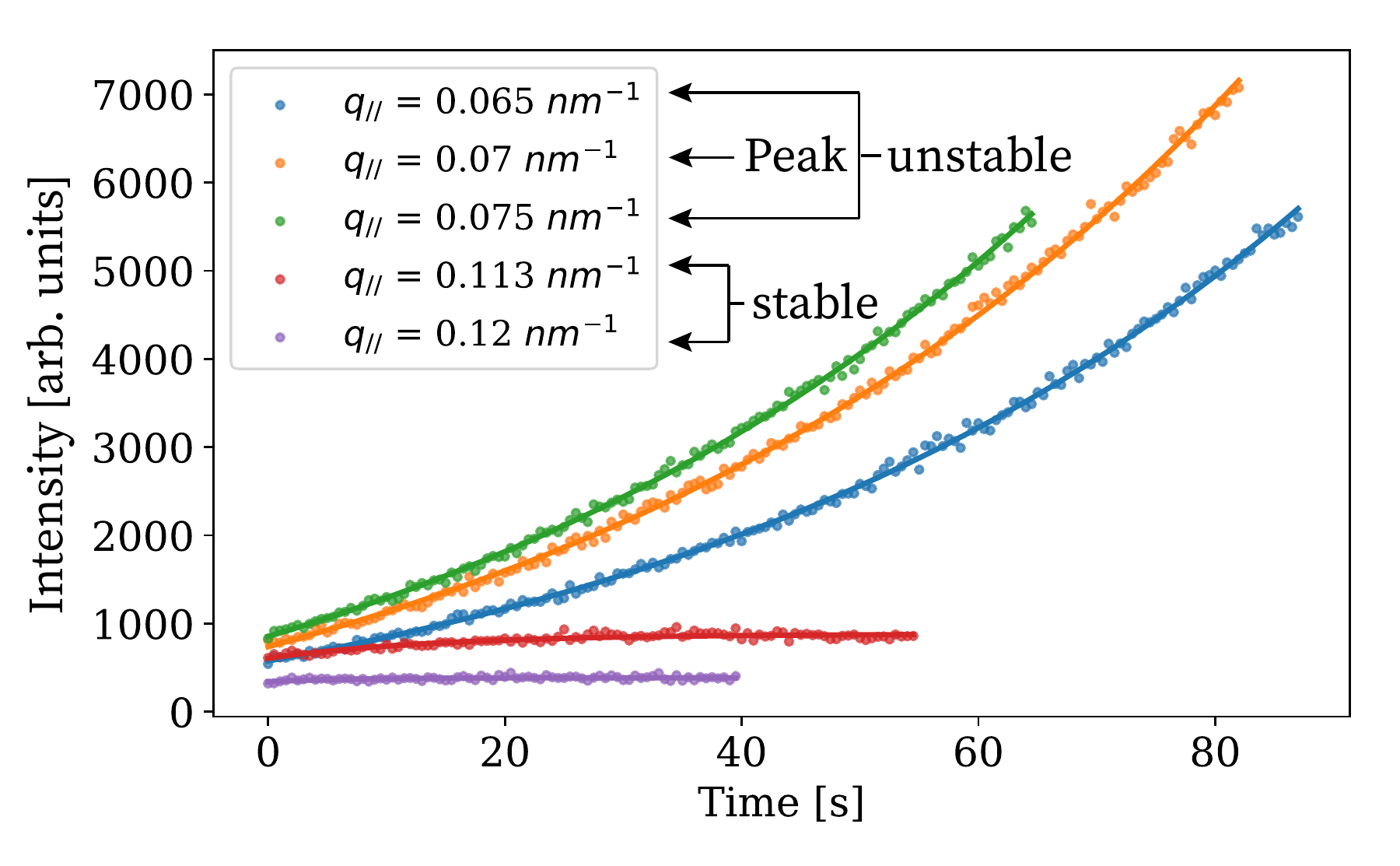}
    \caption{Linear theory fits (solid lines) to intensity evolution as a function of time for different wavenumbers in the high temperature crystalline nanoscale patterning.}
    \label{fig:I_t}
\end{figure}

To determine $R(\mathbf{q})$, the $\mathbf{q}$-averaged intensities were first computed by averaging 10 detector pixels in the $q_{||}$ direction ($\Delta q_{||}$ = $\pm$0.004 nm$^{-1}$) and 70 pixels in the $q_z$ direction ($\Delta q_{z}$ = $\pm$0.028 nm$^{-1}$). The temporal evolution of the scattered intensity from each wavenumber bin was then fitted with a function of the form $I(q_{||},t) = a  e^{2Rt} + b$, as in Eq. \ref{equ:hhstructure-factor}, with $a$, $b$ and $R$ being independent fit parameters for each $q_{||}$ bin. The time scales over which the linear theory is valid differ from one $q_{||}$ bin to another since structural evolution happens faster at smaller length scales. In Fig. \ref{fig:I_t}, the intensity fits are shown for selected $q_{||}$ in high temperature patterning. For each wavenumber of interest, $\chi^2$ was computed to determine optimum time scales over which the fits were performed. Near the peak wavenumber, there is an exponential growth, indicating surface instability.  On the other hand, the measured $R(q_{||})$ for room temperature bombardment are all negative, showing surface stability at all wavenumbers.

The amplification factor is usually taken to have a form\cite{madi2011mass}: 
\begin{equation}
R(q_{||})=-S_{y}\,q_{||}^2-B\,q_{||}^4
\label{equ:long-wave}
\end{equation}
where $S_y$ is a coefficient of curvature-dependent surface evolution, incorporating the curvature-dependent sputter erosion (the $\nu$ term in Eq. \ref{equ:Oucontinuum}), lateral mass redistribution (the $\nabla \bm{j}_{\mathrm{ion}}$ term), and, in the case of a crystalline surface, the linear part of the ES term $\epsilon$.  The coefficient $B$ (the $\kappa$ term in the case of high temperature patterning according to Eq. \ref{equ:Oucontinuum}) is a measure of surface diffusion and/or surface confined viscous flow. 

\begin{figure*}
    \centering
    \includegraphics[width=3.2in]{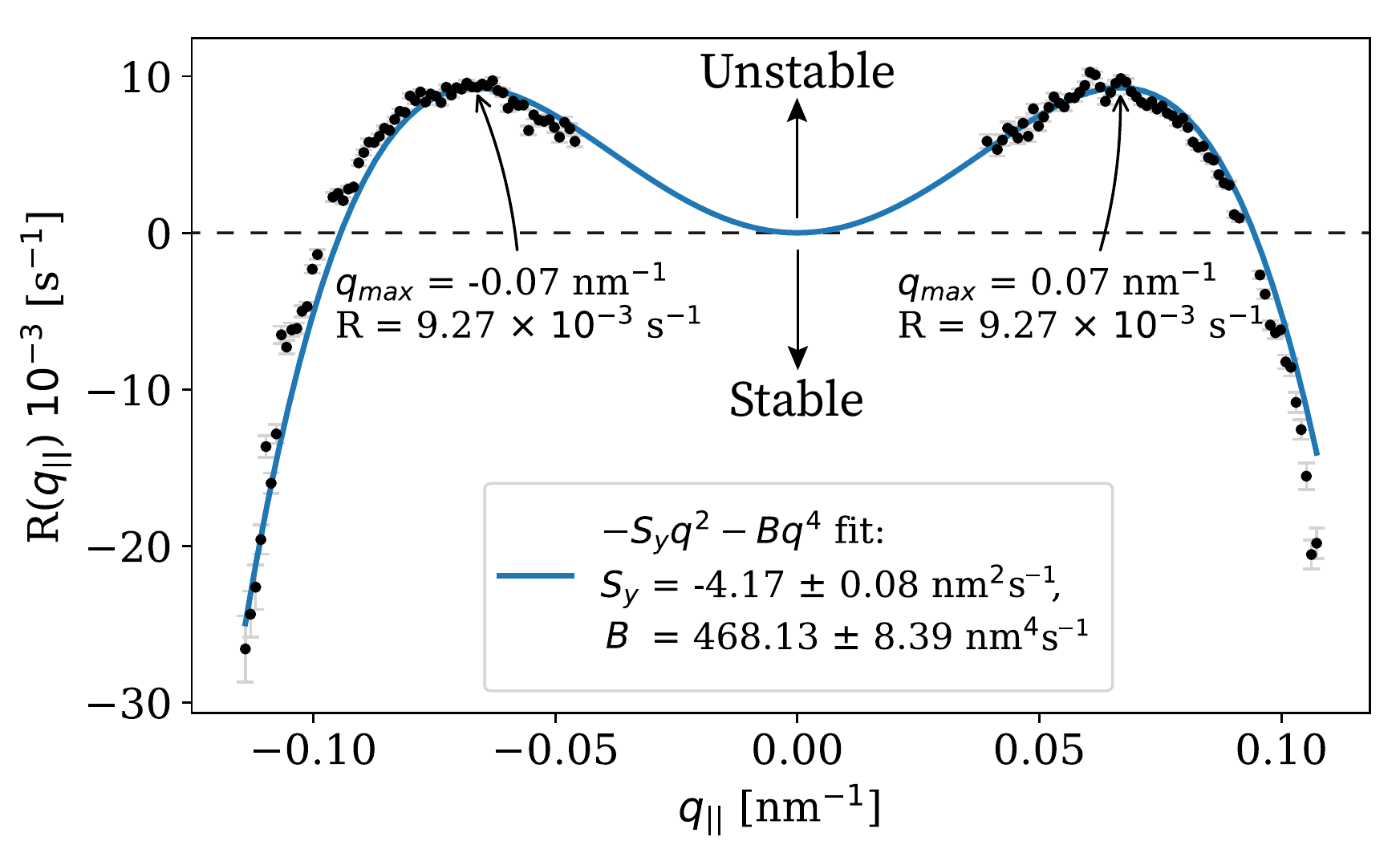}
    \includegraphics[width=3.2in]{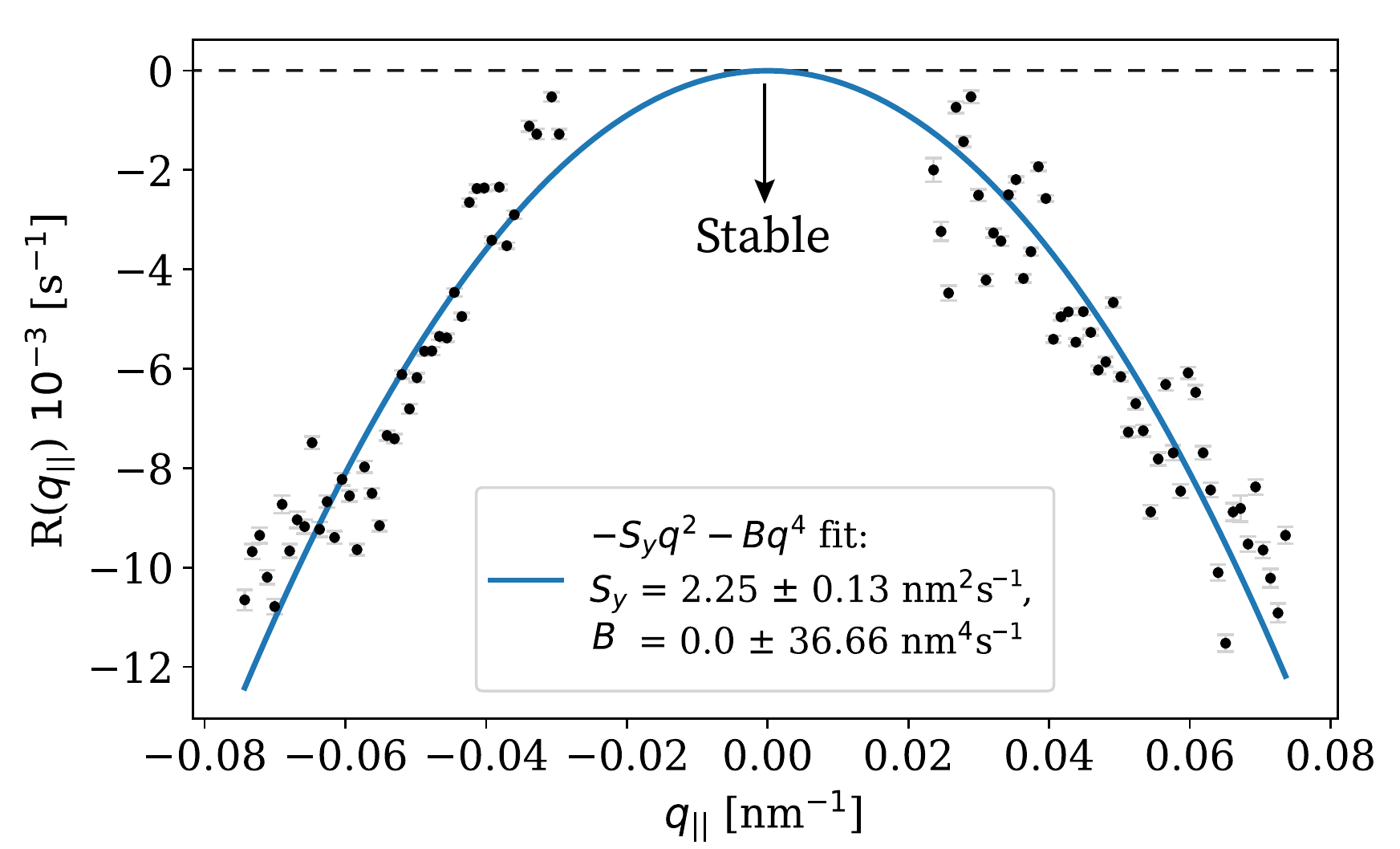}
    \put(-340,140){\textbf{a)}}
    \put(-110,140){\textbf{b)}}
    \caption{a) $R(q_{||})$ measured for the high temperature patterning. b) $R(q_{||})$ measured for the room temperature smoothing. Solid lines are fits to Eq. \ref{equ:long-wave}. Values of $R(q_{||})$ at low wavenumbers have been omitted because those wavenumbers overlap the tails of the reflected beam.}
    \label{fig:R}
\end{figure*}

Figure \ref{fig:R}(a) shows that the R($q_{||}$) values in the high temperature experiment were well fitted with Eq. \ref{equ:long-wave}. The values at the lowest $\left|q_{||}\right|$ were excluded because of the overlap of the reflected beam with the diffuse scattering. Additionally, the behavior is symmetric between positive and negative $q_{||}$ values. The fitted values are $S_y^{HT}$ = -4.17 $\pm$ 0.08 nm$^2$s$^{-1}$ and $B^{HT}$ = 468.1 $\pm$ 8.4 nm$^4$s$^{-1}$. In linear theory, the wavenumber of most rapid growth, $q^{max}_{||}$, is: $q^{max}_{||} = \sqrt{|S_y|/(2B)} = 0.07 \; \mathrm{nm}^{-1}$. In other words, the developing structures had an initial average length scale of $2\pi / 0.07 \; \mathrm{nm}^{-1} \; \sim$ 90 nm, which is comparable to the length scales in Fig. \ref{fig:AFM}, though some coarsening had taken place by the time of the \textit{post-facto} image, resulting in a slightly larger period of the patterns. 

For the room temperature smoothing experiment, the pre-patterned Ge sample was bombarded beforehand and had well-developed faceted patterns with four-fold symmetry, similar to the final result of the high-temperature experiment as shown in Fig. \ref{fig:AFM}. After 10 seconds of bombardment, amorphization had occurred and decrease in x-ray intensity indicates that the large nanoscale structures had been significantly suppressed. Then, the scattering intensities from the pre-patterned sample started to decay at exponential rates, which were well fitted by the linear theory model, yielding $R(q_{||})$. The measured $R(q_{||})$ values in the room temperature experiment were then fitted with Eq. \ref{equ:long-wave} as shown in Fig. \ref{fig:R}(b). 

The results for the two samples are tabulated in Table \ref{tab:linear-th}. The room temperature curvature-dependent term $S_y^{RT}$ = 2.25 $\pm$ 0.13 nm$^2$s$^{-1}$ has opposite sign compared to $S_y^{HT}$. This difference in sign is expected due to the instability in the high temperature experiment leading to the formation of patterns versus the surface stability and smoothing in the room temperature experiment.

As noted above, the $S_y$ coefficient in Eq. \ref{equ:long-wave} includes curvature dependent sputtering $\nu$, lateral mass redistribution, in the high temperature patterning case, it also includes surface instability due to the presence of the ES barrier term. Since the incident ion and most recoil energies are far above thermal energies, we expect that contributions from sputter erosion and lateral mass redistribution are relatively independent of temperature, even through the change in bombarded material from amorphous to crystalline with increasing temperature.   However the ES barrier term only exists in the high temperature patterning case. Therefore, subtraction of the curvature-dependent term measured at the room temperature smoothing $S_y^{RT}$ from the curvature-dependent term at high temperature patterning $S_y^{HT}$ gives the ES contribution to the curvature-dependent term: $\epsilon$ = $S_y^{HT}-S_y^{RT}$ = -6.42 $\pm$ 0.15 nm$^2s^{-1}$. 

It is worth noting that the erosive and redistributive contributions to $S_y^{RT}$ can be estimated following the general approaches of Bobes \textit{et al.} \cite{bobes2012ion} and Hofs{\"a}ss \cite{hofsass2014simulation} using SDTrimSP \cite{mutzke2019sdtrimsp} binary collision approximation simulations in conjunction with the erosive formalism of Bradley and Harper \cite{bradley1988theory} and the redistributive formulism of Carter and Vishnyakov \cite{carter1996roughening}. These give an erosive contribution $S_y^{eros} \approx -0.29$ nm$^2$/s and a redistributive contribution $S_y^{redist} \approx 0.68$ nm$^2$/s, for a total $S_y^{eros+redist} \approx 0.39$ nm$^2$/s. A different approach to calculating the curvature coefficient \cite{norris2014pycraters} using the PyCraters Python framework \cite{PyCraters2017} for crater function analysis on the SDTrimSP results also gives $S_y^{total} \approx 0.40$ nm$^2$/s. Both of these values are significantly smaller than the measured value of 2.25 nm$^2$/s, but estimates of the parameters entering theory must be considered approximate. 

Turning to the smoothening term $B$, high wavenumbers play an important role since it multiplies $q_{||}^{4}$, unlike the case for $S_y$ which multiplies $q_{||}^{2}$ and can be determined reliably by $R(q_{||})$ values at low wavenumbers. The large uncertainty in the experimental value of $B^{RT}$ can therefore be explained by the fact that smoothing causes limited scattering intensities at high wavenumbers. In contrast, $B^{HT}$ has a relatively small uncertainty due to a wider range of accessible wavenumbers. 

In the room temperature experiment, $B^{RT}$ is believed to be due primarily to ion-induced surface viscous flow\cite{umbach2001spontaneous}.  It is generally believed that ion-induced diffusion and any thermal relaxation are much smaller at room temperature. Because the surface remains crystalline during high temperature nanopatterning $B^{HT}$, on the other hand, is presumably due to surface diffusion, which could  be a combination of thermal and/or ion-induced.  As far as we know, there is no compelling rationale suggesting that a pure ion-induced surface diffusion should be much larger on a crystalline surface than on an amorphous surface.  Therefore it appears that the large value of $B^{HT}$ relative to $B^{RT}$ is because of thermal relaxation processes.

\begin{table}
\caption{\label{tab:linear-th}%
Comparative linear theory analysis results}
\begin{ruledtabular}
\begin{tabular}{cccc}
$T_{\mathrm{sample}}$ &
\multicolumn{1}{c}{\textrm{$S_y$}}&
\multicolumn{1}{c}{\textrm{$B$}}&
\multicolumn{1}{c}{\textrm{$q_{max}$}}\\
 & [nm$^2s^{-1}$] & [nm$^4s^{-1}$] & [nm$^{-1}$] \vspace{1mm}\\
\hline 
RT\::\;\;30$^\circ$C&\:\,2.25 $\pm$ 0.13&\;\;\;\;\,0.0 $\pm$ 36.7&--- \\
HT:\,300$^\circ$C&-4.17 $\pm$ 0.08&\;468.1 $\pm$ \;\,8.4&0.07 \\
\hline
\hline 
\, &
\multicolumn{1}{c}{\textrm{$\epsilon$}}&
\multicolumn{1}{c}{\,}&
\,\\
 & $\left(S_y^{HT}-S_y^{RT}\right)$ & \, & \, \vspace{1mm}\\
 \hline
 Subtraction &-6.42 $\pm$ 0.15& \, &\, \\
\end{tabular}
\end{ruledtabular}
\end{table}
In summary, the ES barrier contribution to the curvature-dependent patterning of Ge was determined by subtracting the measured curvature-dependent term for room temperature smoothing from that of high temperature patterning. Compared to the erosive and lateral mass redistribution contributions to the curvature-dependent term, the magnitude of the ES barrier term is considerably larger in magnitude. Within the context of the continuum equation Eq. \ref{equ:Oucontinuum}, the ES barrier contribution $\epsilon$ is primarily responsible for the instability of the crystalline Ge surface. Development of a theoretical formalism linking the measured ES barrier kinetic term to the size of the ES barrier energy in the self-organized ion beam nanopatterning would be a very interesting direction of future research and allow comparison with the existing ES barrier measurement in Ge(001) homoepitaxy\cite{shin2007modeling}.

This material is based on work partly supported at BU by the National Science Foundation (NSF) under Grant No. DMR-1709380. X.Z. and R.H. were supported by the U.S. Department of Energy (DOE) Office of Science under Grant No. DE-SC0017802. Experiments were performed at the Coherent Hard X-ray (CHX) beamline at National Synchrotron Light Source II (NSLS-II), a U.S. Department of Energy (DOE) Office of Science User Facility operated for the DOE Office of Science by Brookhaven National Laboratory under Contract No. DE-SC0012704. We thank Glenn Thayer and Heitor Mourato of the Boston University Scientific Instrumentation Facility for design and construction of the sample holder.

\bibliography{ionpatterning_references.bib}

\end{document}